%% file: final_paper_v1.tex
\documentclass[sigconf]{acmart}

\AtBeginDocument{%
  \providecommand\BibTeX{{%
    \normalfont B\kern-0.5em{\scshape i\kern-0.25em b}\kern-0.8em\TeX}}}

\usepackage[ruled,vlined]{algorithm2e}

\usepackage{multirow}
\usepackage{dsfont}
\usepackage{balance}

\copyrightyear{2023}
\acmYear{2023}
\setcopyright{acmlicensed}\acmConference[CIKM '23]{Proceedings of the 32nd ACM International Conference on Information and Knowledge Management}{October 21--25, 2023}{Birmingham, United Kingdom}
\acmBooktitle{Proceedings of the 32nd ACM International Conference on Information and Knowledge Management (CIKM '23), October 21--25, 2023, Birmingham, United Kingdom}
\acmPrice{15.00}
\acmDOI{10.1145/3583780.3614657}
\acmISBN{979-8-4007-0124-5/23/10}

\begin{document}

\title{An Unified Search and Recommendation Foundation Model for Cold-Start Scenario}

\author{Yuqi Gong}
\affiliation{%
  \institution{Ant Group}
  \city{Beijing}
  \country{China}
  }
\email{gongyuqi.gyq@antgroup.com}

\author{Xichen Ding}
\affiliation{%
  \institution{Ant Group}
    \city{Beijing}
  \country{China}
}
\email{xichen.dxc@antgroup.com}

\author{Yehui Su}
\affiliation{%
  \institution{Ant Group}
    \city{Beijing}
  \country{China}
}
\email{suyehui.syh@antgroup.com}

\author{Kaiming Shen}
\affiliation{%
  \institution{Ant Group}
    \city{Beijing}
  \country{China}
}
\email{kaiming.skm@antgroup.com}

\author{Zhongyi Liu}
\affiliation{%
  \institution{Ant Group}
      \city{Hangzhou}
  \country{China}
}
\email{zhongyi.lzy@antgroup.com}

\author{Guannan Zhang}
\affiliation{%
  \institution{Ant Group}
    \city{Hangzhou}
  \country{China}
}
\email{zgn138592@antgroup.com}

\renewcommand{\shortauthors}{Yuqi Gong et al.}


\input{1abstract}







\input{1abstract}
\maketitle

\input{2intro}
\input{4model}
\input{5exp}
\input{6conclusion}

\bibliographystyle{ACM-Reference-Format}
\balance
\bibliography{final_paper_v1}

\end{document}

%% file: 1abstract.tex
\begin{abstract}

In modern commercial search engines and recommendation systems, data from multiple domains is available to jointly train the multi-domain model. Traditional methods train multi-domain models in the multi-task setting, with shared parameters to learn the similarity of multiple tasks, and task-specific parameters to learn the divergence of features, labels, and sample distributions of individual tasks. With the development of large language models, LLM can extract global domain-invariant text features that serve both search and recommendation tasks. We propose a novel framework called S\&R Multi-Domain Foundation, which uses LLM to extract domain invariant features, and Aspect Gating Fusion to merge the ID feature, domain invariant text features and task-specific heterogeneous sparse features to obtain the representations of query and item. Additionally, samples from multiple search and recommendation scenarios are trained jointly with Domain Adaptive Multi-Task module to obtain the multi-domain foundation model. We apply the S\&R Multi-Domain foundation model to cold start scenarios in the pretrain-finetune manner, which achieves better performance than other SOTA transfer learning methods. The S\&R Multi-Domain Foundation model has been successfully deployed in Alipay Mobile Application's online services, such as content query recommendation and service card recommendation, etc.

\end{abstract}

%% file: 2intro.tex
\vspace{-1.0em}
\section{Introduction}

Modern commercial recommendation systems and search engines are widely used in online service platforms such as YouTube, TikTok, Taobao, Alipay, etc. Search and recommendation can facilitate users' behaviors to browse instant videos, buy products, use services, and make payments using E-wallets. The similarity between Search (S) and Recommendation (R) makes jointly modeling S\&R a promising research topic. Some work \cite{kuai_sr,causal_learning_recommendation_using_search} propose to enhance recommendation by learning from a unified sequence of search and recommendation behaviors. Others \cite{multi_interest_product_search,hierarchical_embeb_product_search} propose to improve personalized search by adding users' multi-interests in recommendation. Methods to model search and recommendation tasks jointly are also proposed in ~\cite{dl_match_sr,DBLP:conf/wsdm/ZamaniC20,DBLP:conf/desires/ZamaniC18,user_unified_sr,e_commerce_sr}. JSR framework in \cite{DBLP:conf/desires/ZamaniC18} simultaneously learns retrieval and recommendation models with shared item set and optimizes a joint loss function. Researchers in \cite{user_unified_sr} applied two-level transformer encoders, text encoder to learn documents and queries, session encoder to model the integrated sequence of search and browsing behaviors. \cite{e_commerce_sr} applied GNN to learn node embedding of user\&item, and treat search query as a special attribute of edges in the graph. In industrial scenarios, there are several benefits to model S and R jointly. First, there are multiple search and recommendation scenarios in a single mobile application. The training data collected in a single domain can't fully reflect users' complete intents and is sub-optimal compared to modeling them jointly. Secondly, majority of items are shared between search and recommendation. Once users have impressions of any products, videos, and services in a recommendation scenario, they should be able to retrieve the item in Search later for repurchase or reuse purposes. Despite the similarity between Search and Recommendation, there are also difficulties in modeling them jointly, such as the data imbalance issue of multiple domains, the heterogeneous issue of different item sets (videos, products, and services), the negative transferring issue. With the latest developments in large language models (LLM) ~\cite{DBLP:conf/nips/BrownMRSKDNSSAA20,DBLP:journals/corr/abs-2303-18223,DBLP:journals/corr/abs-2302-13971,DBLP:journals/corr/abs-2205-08084,cui2022m6rec}, the pretrain-finetune framework ~\cite{e_commerce_sr,kuai_sr,user_unified_sr,infer_implicit,cikm_prototipical_contrastive} greatly improves the performance of downstream tasks. We are inspired by the strong expressive power of natural language features and propose to build the Search and Recommendation Foundation model on top of LLMs, which extract low-level domain-invariant text features of the query (Q) and item (I). The major difference between our S\&R foundation model and traditional multi-domain multi-task models is how we use the domain-invariant text features to help constrain the divergence of different tasks, which alleviates data imbalance, negative transferring, and item heterogeneous issues.
To summarize, our proposed S\&R Foundation model has the following key contributions: 
\begin{itemize}
\item{
We apply LLMs in S\&R Multi-Domain Foundation model, and extract domain invariant text features to help mitigate the negative transferring and item heterogeneous issues in the multi-domain settings.
}
\item{
We novelly proposed the Aspect Gating Fusion (Domain-Specific Gating) to fuse the ID feature, text features from LLMs, and sparse features. The Domain Adaptive Multi-Task module is also used to extract the domain-specific query and item towers' representations. 
}
\item{
For the cold start of new scenarios, we have conducted extensive experiments both offline and online, to show the effectiveness of supervised fine-tuning of our S\&R Foundation model in downstream tasks, which is now fully deployed online and serving in Alipay's mobile application.	
}
\end{itemize}

%% file: 4model.tex
\vspace{-1.0em}

\begin{figure*}
  \centering
  \includegraphics[height=3.5in, width=4.9in]{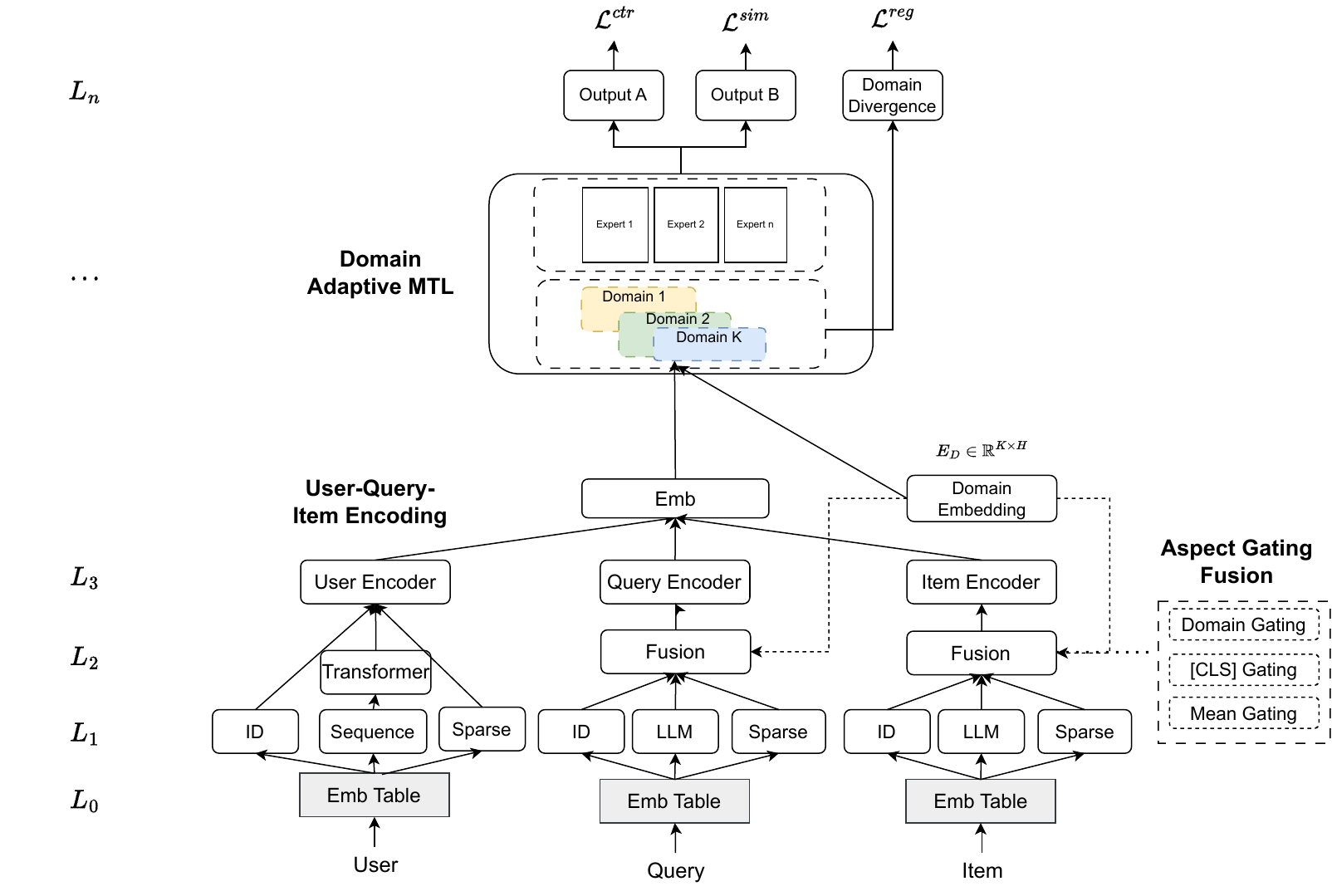}
  \caption{SR Multi-Domain Foundation Model Architecture}
  \label{fig:foundation}
\end{figure*}

\section{Proposed Model}
\subsection{Problem Formulation}
Given a set of $K$ search and recommendation tasks $\{D_{k}\}^{K}_{k=1}$, $D_{k}$ denotes the dataset for the $k$-th task. We let $\mathcal{U}=\{u_{1},u_{2},…,u_{N}\}$ denote the user set, $\mathcal{I}=\{i_{1},i_{2},…,i_{M}\}$ denote the item set and $\mathcal{Q}=\{q_{1}, q_{2}, ..., q_{T}\}$ denote the search query set. In real-world scenarios, items in search and recommendation usually come from different domains and are heterogeneous. Some items are shared across multiple domains and some items belong to each specific domain. And we let $\mathcal{I}=\mathcal{I}_{1} \cup \mathcal{I}_{2} \cup … \cup \mathcal{I}_{K}$ denote the union of all items in $K$ domains, which contains $M$ items in total. We aim to jointly train a search and recommendation (S\&R) foundation model $M^{S\&R}_{Foundation}$ in the multi-task setting and predict the probability of user $u_{l}$ click the item $i_{l}$ given input query $q_{l}$ as $p(y^{ctr}_{l}=1|u_{l},q_{l}, i_{l})$. And for search scenarios, additional query-item relevance score is also predicted as $p(y^{sim}_{l}=1|q_{l},i_{l})$. 
For cold start of a new search or recommendation scenario $D^{*}$, we restore parameters of embedding tables and partial network structures from the pretrained S\&R foundation model $M^{S\&R}_{Foundation}$, and then apply supervised fine-tuning on the downstream tasks, such as click through rate (CTR) prediction, query-item relevance prediction, etc. For the search task $D^{S}_{k}$, we let $D^{S}_{k}=\{x_{l}=(u_{l},q_{l},i_{l}), y_{l}\}_{l}$, which denotes the search ranking task given the triple input of (user, query, item) as $(u_{l},q_{l},i_{l})$. For the recommendation task $D^{R}_{k}$, we set search query set Q as emptyset $\emptyset$ in $D^{R}_{k}=\{x_{l}=(u_{l},q_{l}=\emptyset,i_{l}), y_{l}\}_{l}$.

\subsection{S\&R Multi-Domain Foundation Model}

As illustrated in Figure \ref{fig:foundation}, the S\&R Multi-Domain Foundation model has three main components: the User-Query-Item encoding module, the Aspect Gating Fusion module, and the Domain-Adaptive Multi-Task module. Firstly, raw features of user, query and item pass through the embedding layers, and we extract the ID embedding, token-level text embedding and sparse features' embedding. We apply LLM to extract domain-invariant text features of query and item towers, which minimize the divergence of features' distribution cross multiple domains. Secondly, the Aspect Gating Fusion module is designed to merge different groups of ID, text, sparse features' embedding. The fusion network is to balance the relative importance of ID, text, and sparse features. Very few training samples contain ID features of cold start items and can't represent them well, and generic text features play more important role. Finally, we feed the concatenated embedding of user, query and item towers to the Domain Adaptive MTL module. The module has two outputs representing the click through rate (CTR) prediction task and the query-item relevance prediction task. The final loss function is the sum of CTR prediction loss $\mathcal{L}^{ctr}$, relevance prediction loss $\mathcal{L}^{sim}$ and domain adaptive regularization $\mathcal{L}^{reg}$.

\vspace{-1.0em}

\subsubsection{User Query and Item Encoding}


We extract three towers for user, query and item respectively. For the user tower, $e_{u}^{ID} \in \mathbb{R}^{D}$ denotes user id embedding. $e_{u}^{N_{H}}=[x_{1},...,x_{s},…,x_{N_{H}}]$ denotes the unified sequence of both search and recommendation clicks in chronological order. Each behavior $x_{s}$ is encoded as multiple layers of MLPs with inputs of ID feature, sparse feature of behavior type S or R, and other sparse features of attributes, $x_{s}=FC(e^{ID}_{s} \oplus e^{type}_{s} \oplus e^{attr}_{s})$. For the query ($Q$) and item ($I$) features, we extract both domain-invariant text features, such as tokens in search query and items' title, and the domain-specific sparse features. The tokens of $Q$ and $I$ go through the same tokenizer and we get the tokenized id sequences as integer tensors $e^{Token}_{q}$ and $e^{Token}_{i}$. 
$e^{Token}_{q} = [e^{1}_{q},e^{2}_{q},…, e^{L_{q}}_{q}] \in \mathbb{R}^{L_{q} \times D}$ denotes the query's tokenized id tensor of length $L_{q}$, and $e^{Token}_{i} = [e^{1}_{i},e^{2}_{i},…, e^{L_{i}}_{i}] \in \mathbb{R}^{L_{i} \times D}$ denotes the item's tokenized id tensor of length $L_{i}$. For ID feature, we also embed the search query as ID feature $e^{ID}_{q} \in \mathbb{R}^{D}$, and item ID as $e^{ID}_{i} \in \mathbb{R}^{D}$. For the sparse features, we embed sparse features of $Q$ as $e^{S}_{q}$ and sparse features of $I$ as $e^{S}_{i}$. Finally, we get the feature groups of query tower as $e_{q}=[e_{q}^{ID},e_{q}^{Token}, e_{q}^{S}]$ and the feature groups of item tower as $e_{i}=[e_{i}^{ID},e_{i}^{Token}, e_{i}^{S}]$. 

\textbf{LLM as Domain-Invariant Feature Extractor}

We apply the pretrained Large Language Model, such as BERT \cite{DBLP:conf/naacl/DevlinCLT19}, GPT \cite{DBLP:conf/nips/BrownMRSKDNSSAA20}, ChatGLM \cite{zeng2022glm,du2022glm}, to extract domain-invariant text features on both query tower and item tower, represented as $\phi_{lm}(Q)=\phi_{lm}(e^{Token}_{q}) \in \mathbb{R}^{L_{q} \times D}$ and $\phi_{lm}(I)=\phi_{lm}(e^{Token}_{i}) \in \mathbb{R}^{L_{i} \times D}$. After mean pooling of the encoding layer, followed by shared linear projection, we get the domain-invariant text representation of query and item as $E_{lm}(Q)=W_{lm} \times\text{MEAN}(\phi_{lm}(e^{Token}_{q})) \in R^{H}, E_{lm}(I)=W_{lm} \times\text{MEAN}(\phi_{lm}(e^{Token}_{i})) \in R^{H}$. $W_{lm} \in \mathbb{R}^{H \times D}$ denotes the linear projection layer shared between query tower and item tower, $H$ denotes the hidden size of the learned representations. The language models' representation is useful for cold start scenarios, especially when we have few training samples to update the ID feature of new item $i^{*}$ and new search query $q^{*}$. We also apply linear projections $W^{q}_{ID},W^{i}_{ID} \in \mathbb{R}^{H \times D}$ to ID feature of query and item tower, and get the ID representation of query and item as $E_{ID}(Q), E_{ID}(I) \in  \mathbb{R}^{H}$. For the sparse features we have separate network (usually multiple layers of MLPs) to encode query and item as $E_{S}(Q), E_{S}(I) \in \mathbb{R}^{H}$. Finally, we get the feature groups of query tower as $[E_{ID}(Q),E_{lm}(Q),E_{S}(Q)]$ and item tower as $[E_{ID}(I),E_{lm}(I),E_{S}(I)]$.

\subsubsection{Aspect Gating Fusion}
After low level networks $L_{0}$ (embedding tables) and $L_{1}$ (feature encoding layers) in Figure \ref{fig:foundation}, we fuse different aspects of query and item as in literature \cite{multi-aspect-dense-retrieval}. Each aspect $E_{a}$ represents some fine-grained properties of query and item, such as ID, text and sparse features. $\mathcal{A}$ denotes the set of aspects we extract from query and item. In S\&R scenarios, we set $|\mathcal{A}|=3$ as ID, text and sparse attributes. Final representations are fused as weighted sum of different aspects' representations. 

$$ E(Q)=\sum_{a}w_a(Q) E_{a}(Q), E(I)=\sum_{a}w_a(I) E_{a}(I)\quad \forall a \in |\mathcal{A}|$$

The weight vector $w(Q), w(I) \in \mathbb{R}^{|\mathcal{A}|}$ are outputs of a gating network, and we have different strategies to design the network. 

\begin{itemize}
\item{\textbf{Mean-Gating Strategy}
Simply mean pooling of different aspects of query and item features as $w_{a}=\frac{1}{|\mathcal{A}|}$. 
}
\item{\textbf{[CLS]-Gating Strategy}
We use randomly initialized embedding $E_{CLS}(Q)$,$ E_{CLS}(I) \in \mathbb{R}^{H}$ to represent classification token [CLS] of query and item respectively. $$w_{a}=\frac{e^{E_{CLS}E_{a}}}{\sum_{a \in |\mathcal{A}|} e^{E_{CLS}E_{a}}} \in \mathbb{R}^{|\mathcal{A}|}$$. 
}

\item{\textbf{Domain-Gating Strategy\label{domain_gating}}

We design the domain gating strategy from the intuition that the fusion network has different weights when merging different aspects of query and item. 
To model the differences across domains, we randomly initialize the domain embedding
$ E_{D}=[E_{D_{1}},E_{D_{2}},...,E_{D_{K}}]\in \mathbb{R}^{K\times H} $ as the representations of different domains. And the domain-specific gating is calculated as $$\\w_{a}=\frac{e^{E_{D_{k}}E_{a}}}{\sum_{a \in |\mathcal{A}|} e^{E_{D_{k}}E_{a}}} \in \mathbb{R}^{|\mathcal{A}|}$$. 

}

\end{itemize}

\vspace{-1.0em}

\subsubsection{Domain Adaptive Multi-Task Learning}

The input to the Domain Adaptive Multi-Task module is the concatenation of representations of user, query and item towers as $\bold{x}=E(U) \oplus E(Q) \oplus E(I)$. For multi-domain setting, a series of multi-task and multi-domain models are proposed, such as SharedBottom\cite{DBLP:journals/ml/Caruana97}, MMoE\cite{DBLP:conf/kdd/MaZYCHC18}, PLE\cite{DBLP:conf/recsys/TangLZG20}, STAR\cite{DBLP:conf/cikm/ShengZZDDLYLZDZ21}, SAMD\cite{samd_multi_scenario}, etc. These models use shared structures (Experts or MLP layers) to model the similarity among different tasks or domains, and use individual structures to learn the domain-specific properties. The difficulty of training the multi-domain models is the domain shift phenomena. For the k-th domain $D_{k}$, the marginal distribution of input feature $p(\bold{x}_{k})$ and the conditional distribution of predicting output $y_{k}$ as $p(y_{k}|\bold{x}_{k})$ has divergence from other domains. The well studied MTL models handle the divergence of conditional distribution. We propose to add a Domain Adaptive Layer to the input features $\bold{x}_{i}$, which maps the inputs from multiple domains to a common vector space. We reuse the randomly initialized domain embedding $E_{D}=[E_{D_{1}},E_{D_{2}},...,E_{D_{K}}]\in \mathbb{R}^{K\times H}$ in section \ref{domain_gating} and concatenate the domain embedding $E_{D_{k}}$ to feature vector $\bold{x}_{i}$ of instances from the k-th domain $D_{k}$, followed by domain-specific linear transformation $W_{k}$. Suppose $\bold{x}_{i}$ and $\bold{x}_{j}$ denote two instances from different domains in the same training batch, we can get the domain-adaptive representation $\hat{\bold{x}}_{i},\hat{\bold{x}}_{j}$ as

$$\hat{\bold{x}}_{i}=W_{i}(\bold{x}_{i} \oplus E_{D_{i}}),\hat{\bold{x}}_{j}=W_{j}(\bold{x}_{j} \oplus E_{D_{j}})$$

We apply domain adaptation \cite{dann_ganin} techniques to constrain the divergence of distributions from domains $i$ and $j$ as $p(\hat{\bold{x}}_{i})$ and $p(\hat{\bold{x}}_{j})$ as $\mathcal{L}^{reg}=\sum_{i,j \in \{1,2,...,K\}}d(p(\hat{\bold{x}}_{i})||p(\hat{\bold{x}}_{j}))$. In terms of divergence measurement, we compared different metrics such as Jensen-Shannon Divergence (symmetric KL Divergence), Maximum Mean Discrepancy (MMD) \cite{gretton_kernel_test_mmd} in the experiment section. And we find the Jensen-Shannon Divergence achieves the best performance as $$\mathcal{L}^{reg}=\sum_{i,j \in \{1,2,...,K\}} \text{JS}(p(\hat{\bold{x}}_{i}) || p(\hat{\bold{x}}_{j}))$$.

Finally, on top of the Domain Adaptive Layer we stack the standard Multi-Task module, such as MMoE to extract outputs and predict two objectives, CTR prediction $y^{ctr}$ and query-item relevance prediction $y^{sim}$. 

\textbf{CTR Prediction}
Click-Through Rate (CTR) Prediction is a common task in both search and recommendation scenarios. We apply a unified scoring function $y^{ctr}_{l}=f_{\theta}(u_{l},q_{l},i_{l})$ in the S\&R foundation framework to predict CTR with the triple inputs of user, item and query as $(u, q, i)$. For search tasks, users have explicit search query $q$. And for recommendation tasks users don't have explicit intentions. So we set $q=\emptyset$ as the default embedding in the unified scoring function.

\textbf{Query-Item Relevance Prediction}
Query-Item Relevance Prediction is a common task in search scenarios, which predicts the relevance score of query-item pair of $(q,i)$ and train a function $y^{sim}_{l}=f_{\phi}(q_{l},i_{l})$ to represent query-item pair's relevance score. The relevance prediction is usually a classification task.

\subsubsection{Loss of S\&R Foundation model}
We train the S\&R foundation model in multi-domain multi-task settings, using datasets from $K$ domains. Each domain calculates either or both of two objectives of CTR prediction $y^{ctr}_{l}=f_{\theta}(u_{l},q_{l},i_{l})$ and relevance prediction $y^{sim}_{l}=f_{\phi}(q_{l},i_{l})$, depending on whether the task is search or recommendation. The final objective function consists of three parts, the loss for CTR prediction $\mathcal{L}^{ctr}$, the loss for relevance prediction $\mathcal{L}^{sim}$, and the loss for domain adaptive regularizer $\mathcal{L}^{reg}$.

$$\mathcal{L}=\mathcal{L}^{ctr}+\mathcal{L}^{sim}+\mathcal{L}^{reg}$$

$$\mathcal{L}^{ctr}=\sum_{k\in K}\sum_{l \in N^{ctr}_{k}}\mathcal{L}_{ce}(f_{\theta}(u_{l},q_{l},i_{l});y^{ctr}_{l})$$

$$\mathcal{L}^{sim}=\sum_{k \in K}\sum_{l \in N^{sim}_{k}}\mathcal{L}_{ce}(f_{\phi}(q_{l},i_{l});y^{sim}_{l})$$

\subsection{Supervised Fine-Tuning Downstream Tasks}
The pretrained S\&R foundation model can benefit downstream tasks in the pretrain-finetune manner. The downstream model restores parameters from the foundation model, freezes part of the parameters and finetunes the remaining layers. We experiment different ways of freeze-finetune split. Firstly, the freeze-finetune split is between level $L_{0}$ and $L_{1}$ as in Figure \ref{fig:foundation}. The pretrained embedding in level $L_{0}$ is freezed and the remaining layers from $L_{1}$ to $L_{n}$ are finetuned. Secondly, the freeze-finetune split is between level $L_{1}$ and $L_{2}$. The embedding in level $L_{0}$ as well as the parameters of encoding layers in level $L_{1}$ are freezed, and the parameters from level $L_{2}$ to $L_{n}$ are finetuned. Given dataset of new downstream task $D^{*}=\{(u^{*}_{l}, q^{*}_{l}, i^{*}_{l}),y^{*}_{l} \}$, the domain embedding $E_{D^{*}} \in \mathbb{R}^{H}$ is randomly initialized and finetuned. In the experiment section, we thoroughly tested the performance of different ways of freeze-finetune split. We also compared the performance of pretrain-finetuning S\&R Foundation model $M^{S\&R}_{Foundation}$ with the performance of training single domain model without transfer learning.


%% file: 5exp.tex
\vspace{-1.0em}

\section{Experiment}
To test the effectiveness of our proposed S\&R Multi-Domain Foundation model, we want to answer the following questions:

\begin{itemize}
\item{\textbf{RQ1:}
Whether our joint S\&R Multi-Domain Foundation model can achieve SOTA performance compared to other multi-domain and multi-task models? 
}
\item{\textbf{RQ2:}
In terms of query and item towers’ representations, what's the performance of the domain-invariant text features extracted by LLM and Aspect Gating Fusion network compared to other methods?
}
\item{\textbf{RQ3:}
Whether S\&R Multi-Domain Foundation and Supervised Finetuning can help benefit cold start scenarios? 
}
\end{itemize}

\begin{table*}
  \setlength{\abovecaptionskip}{0cm}  
  \setlength{\belowcaptionskip}{-0.1cm} 
  \caption{Statistics of Alipay Search Ranking and Query Recommendation Datasets.}
  \label{tab:statistics}
  \begin{tabular}{|c|c|c|c|c|c|c|c|}
    \hline
    ID & Dataset & S/R & Train & Eval & Test & \#Query & \#Item \\
    \hline
    Task 1  & Query-Item Relevance Prediction & S & 76.2M & 12.7M & 12.7M & 40K & 40K \\
    \hline
    Task 2 & Query Suggest CTR Prediction & S/R & 145.4M & 23.5M & 23.5M & 0.84M & 0.16M \\
    \hline
    Task 3 & Background Word Query Recommendation CTR Prediction & R & 146.2M & 24.3M  & 24.3M & - & 65K \\
    \hline
    Task 4 & Content Query Recommendation CTR Prediction & R & 0.76M & 0.09M & 0.09M & - & 4.6K \\
    \hline
    Task 5 & People Also Ask DeepSuggest & S/R & 2.4M & 0.38M & 0.38M & 0.41M & 25K \\
    \hline
    Task 6 & Service Card Recommendation & S/R & 1.01M & 0.17M & 0.17M & 1.3K & 1.6K \\
    \hline
    Task 7 & Content Search Ranking & S & 6.13M & 1.03M & 1.03M & 0.27M & 0.14M \\
    \hline
\end{tabular}
\end{table*}

\begin{table*}
  \setlength{\abovecaptionskip}{0cm}  
  \setlength{\belowcaptionskip}{-0.1cm} 
  \caption{Performance of S\&R Multi-Domain Foundation Model.}
  \label{tab:multi_doamin_performance}
  \begin{tabular}{|c|c|c|c|c|c|c|c|}
    \hline
    Method & Task1 & Task2 & Task3 & Task4 & Task5 & Task6 & Task7 \\
    \hline
    S\&R Multi-Domain Shared Bottom~\cite{DBLP:journals/ml/Caruana97} & 0.6483 & 0.8993 & 0.7829 & 0.6575 & 0.8511 & 0.8015 & 0.8561 \\
    \hline
    S\&R Multi-Domain MMoE~\cite{DBLP:conf/kdd/MaZYCHC18} & 0.6482 & *0.9003 & 0.7812 & 0.6650 & 0.8463 & 0.7942 & 0.8599 \\
    \hline
    S\&R Multi-Domain PLE~\cite{DBLP:conf/recsys/TangLZG20} & *0.7006 & 0.8981 & 0.7815 & 0.6682 & 0.8487 & 0.7978 & 0.8620 \\
    \hline
    S\&R Multi-Domain Foundation (MLP)  & 0.6827 & 0.8974 & 0.7784 & 0.6683 & 0.8462 & 0.7926 & 0.8629  \\
    \hline
    S\&R Multi-Domain Foundation-MMoE-DA-MMD  & 0.6874 & 0.8942 & *0.7971 & 0.6793 & 0.8564 & 0.8203 & 0.8569  \\
    \hline
    S\&R Multi-Domain Foundation-MMoE-DA-JS & 0.6942 & 0.8973 & 0.7912 & *0.6979 & *0.8703 & *0.8312 & *0.8692  \\
    \hline
    Absolute Improvement & +0.0459 & -0.0020 & +0.0083 & +0.0404 & +0.0192 & +0.0297 & +0.0131 \\
    \hline
\end{tabular}
\end{table*}

\begin{table*}
  \setlength{\abovecaptionskip}{0cm}  
  \setlength{\belowcaptionskip}{-0.1cm} 
  \caption{Comparison of Query and Item Token Encoding Methods after Fine-tuning Task 4. }
  \label{tab:query_item_encoder}
  \begin{tabular}{|c|c|c|c|c|}
    \hline
    ID & Token Embedding & Query/Item Encoder & Finetune & AUC \\
    \hline
    1 & Baseline: Without Token Emb & - & - & 0.7524 \\
    \hline
    2 & Randomly Initialized & Mean Pooling & - & 0.7551 \\
    \hline
    3 & Randomly Initialized &  Transfomer(L=1) & True & 0.7544 \\
    \hline
    4 & Randomly Initialized & Transfomer(L=6) & True & 0.7559 \\
    \hline
    5 & SR Foundation (LM=Transformer) & Transfomer(L=1) & $L_{0},L_{1}$:True & 0.7562 \\
    \hline
    6 & SR Foundation (LM=Transformer) & Transfomer(L=1) & $L_{0}$:False,$L_{1}$:True & 0.7531 \\
    \hline
    7 & SR Foundation (LM=Transformer) & Transfomer(L=1) & $L_{0},L_{1}$:False & 0.7574 \\
    \hline
    8 & SR Foundation (LM=BERT) & BERT BASE(L=12) & True & 0.7563 \\
    \hline
    9 & SR Foundation (LM=BERT) & BERT BASE(L=12)& False & *0.7580 \\
    \hline
    10 & SR Foundation (LM=ChatGLM 6B) & ChatGLM 6B Pretrained LLM \cite{zeng2022glm,du2022glm} & False & 0.7518 \\
    \hline
    11 & SR Foundation (LM=ChatGLM 6B) & ChatGLM 6B Pretrained LLM \cite{zeng2022glm,du2022glm} + prompt & False & 0.7503 \\
    \hline
    12 & SR Foundation (LM=ChatGLM2 6B) & ChatGLM2 Pretrained LLM \cite{zeng2022glm,du2022glm} & False & 0.7502 \\
    \hline
    & Absolute Improvement & - &- & +0.0056 \\
    \hline
\end{tabular}
\end{table*}

\begin{table}
  \setlength{\abovecaptionskip}{0cm}  
  \setlength{\belowcaptionskip}{-0.1cm} 
  \caption{Comparison of Aspect Gating Fusion on Task 4.}
  \label{tab:aspect_gating_fusion}
  \begin{tabular}{|c|c|c|}
    \hline
    Method & AUC & Absolute Gain\\
    \hline
    Mean-Pooling & 0.7385 & -  \\
    \hline
    [CLS]-Gating & 0.7515 & +0.0130 \\
    \hline
    Domain-Gating & *0.7524  & +0.0139 \\
    \hline
\end{tabular}
\end{table}

\begin{table}
  \setlength{\abovecaptionskip}{0cm}  
  \setlength{\belowcaptionskip}{-0.1cm} 
  \caption{Comparison of Cold Start Scenarios Task 4 and 6.}
  \label{tab:supervise_finetune}
  \begin{tabular}{|c|c|c|}
    \hline
    & Service Card Rec & Content Query Rec \\
    \hline
    Single Domain & 0.8229 & 0.7295 \\
    \hline
    SR Fdt->Finetune  & 0.8446 & 0.7574 \\
    \hline
    Absolute Improvement & +0.0216 & +0.0279 \\
    \hline
\end{tabular}
\end{table}

\begin{figure*}
  \centering
  \includegraphics[height=1.5in, width=4.725in]{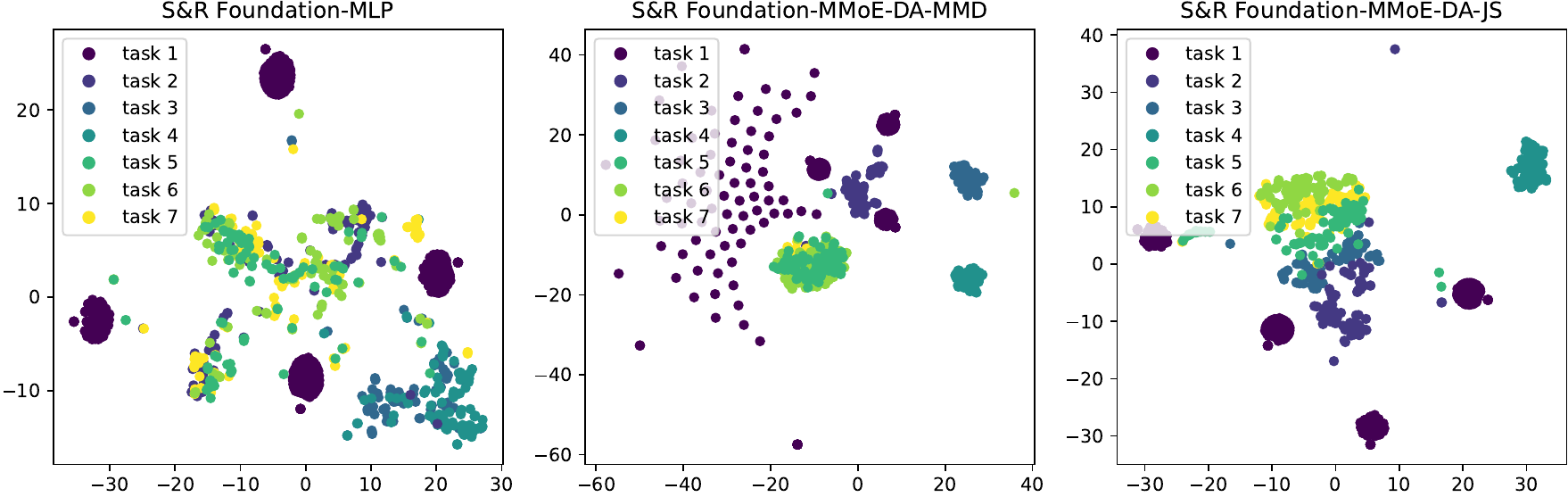}
  \caption{Visualization of SR Foundation Model's Domain-Adaptive Layers}
  \label{fig:domain_adaptive_embedding}
\end{figure*}

\begin{figure}
  \centering
  \includegraphics[height=2.0in, width=2.66in]{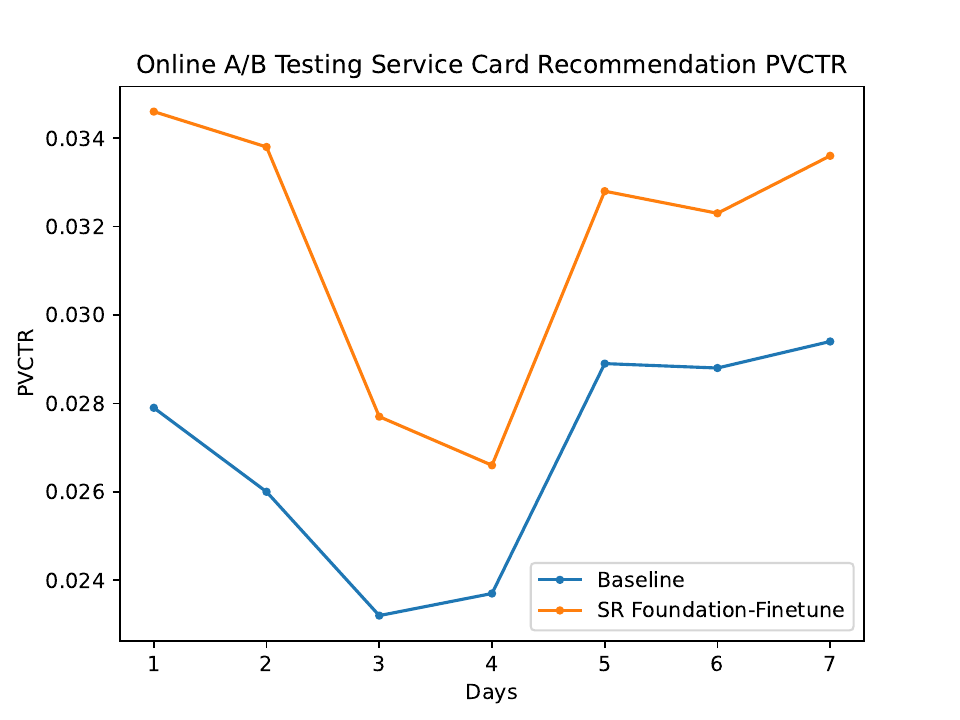}
  \caption{Online AB Testing PVCTR Performance of Service Card Recommendation.}
  \label{fig:service_card_ab_testing}
\end{figure}

\vspace{-1.0em}

\subsection{Experimental Settings}
\subsubsection{Dataset} We conducted extensive experiments of S\&R Foundation model on real-world datasets, including 7 industrial datasets of Alipay Search Ranking and Query Recommendation. The statistics are summarized in table \ref{tab:statistics}. $S$ denotes the search dataset, in which users have explicit search query, such as Query-Item Relevance Prediction, Content Search Ranking, etc. And $R$ denotes the recommendation dataset, in which users don't have explicit intent of search query. There are also some tasks between Search and Recommendation, which we classify as S/R, such as Query Suggest CTR Prediction, in which users have explicit query, and at the same time the task is a CTR prediction task to make recommendation of query suggestions to users.

\subsubsection{Comparison Methods}
\quad \textbf{S\&R Foundation Model} We compared our proposed S\&R Multi-Domain Foundation model with SOTA multi-domain and multi-task models, such as Shared Bottom MTL \cite{DBLP:journals/ml/Caruana97}, Multi-Gate Mixture of Experts (MMoE) \cite{DBLP:conf/kdd/MaZYCHC18}, PLE \cite{DBLP:conf/recsys/TangLZG20}, etc. For ablation study, we designed separate experiments to evaluate different modules of the framework, including the User-Query-Item encoding module, Aspect Gating Fusion module and Domain Adaptive Multi Task module. The experiment of S\&R Multi-Domain Foundation (MLP) denotes the concatenated user-query-item representations are followed by multiple MLP layers. And the experiment of S\&R Multi-Domain Foundation-MMoE-DA-JS denotes the representations are followed by a Domain Adaptive Layer (JS-Divergence) and MMoE multi-task module.

\quad \textbf{Domain-Invariant Text Features and Aspect Gating Fusion}
To prove the effectiveness of adding domain-invariant text features in S\&R Foundation model, we have conducted experiments and ablation studies on different query and item token encoding methods on Alipay Content Query Recommendation dataset of tasks 4 in table \ref{tab:statistics}. In the baseline method, we intentionally leave out the token-embedding of text features and only use ID and sparse features. We also compared randomly initialized token embedding with embedding restored from the pretrained S\&R foundation model under different configurations. For the encoders, we compared mean pooling, randomly initialized Transformer, BERT, ChatGLM-6B\cite{zeng2022glm,du2022glm}, and ChatGLM+prompt, etc. In methods 10-12, we adopt ChatGLM-6B and ChatGLM2-6B to encode the text features of query and items. The implementation details are: we utilized the encoders of ChatGLM-6B and ChatGLM2-6B to convert the input text features and corresponding prompts into 4096-dimensional vectors, which are followed by 2 MLP dense layers and further reduced to 32-dimensional vectors. The prompt function used in our approach is defined as $f_{prompt}(X) = "\text{Extract keywords from sentence [X]}"$. To compare different Aspect Gating Fusion methods, e.g. Mean-Pooling, [CLS]-Gating, Domain-Gating, we conducted ablation studies and the results are listed in table \ref{tab:aspect_gating_fusion}.

\quad \textbf{Supervised Fine-Tuning on Cold Start Scenarios}
For cold start scenarios, we compared the performance of supervised fine-tuning (SFT) the foundation model using downstream dataset, with the method of training single domain model on several tasks, including Service Card Recommendation (task 6) and Content Query Recommendation (task 4) as in table \ref{tab:supervise_finetune}.

\vspace{-1.0 em}

\subsection{Experimental Results}
\subsubsection{S\&R Multi-Domain Foundation model}
To compare the performance of different multi-domain models, we report AUC performance on 7 search and recommendation datasets in table ~\ref{tab:multi_doamin_performance}. All the experimented models share same input features, User-Query-Item Encoding module, and Domain-Gating as Aspect Gating Fusion strategy. The baseline for the multi-task learning (MTL) module is the shared bottom model. MMoE-DA-MMD and MMoE-DA-JS represent models that utilize Maximum Mean Discrepancy (MMD) and Jensen–Shannon Divergence (JS-Divergence) to constrain the distributions of domain adaptive layers respectively. The asterisk (*) denotes the best performance achieved in each task, and the absolute improvement represents the absolute improvement of MMoE-DA-JS method compared to baseline. MMoE-DA-JS achieved best performance on 4 tasks: 4, 5, 6, 7 with AUC improvement of +0.0404, +0.0192, +0.0297, +0.0131 respectively. The domain-adaptive layer constrains the embedding representations from different domains in the common vector space. The t-SNE visualization of S\&R Foundation model's domain-adaptive layers is depicted in Figure \ref{fig:domain_adaptive_embedding}. The embedding depicted in the first subplot "S\&R Foundation MLP" is scattered, and the embedding in the third subplot "S\&R Foundation-MMoE-DA-JS" is coherently aligned.

\vspace{-0.8em}

\subsubsection{Domain-Invariant Text Features and Aspect Gating Fusion}
We report the performance of different methods to encode domain-invariant text features and freeze-finetune split in table \ref{tab:query_item_encoder} on task 4 Content Query Recommendation. Our proposed method of restoring pretrained parameters from BERT BASE (12 layers Transformer) in S\&R Foundation, freezing the parameters of the encoder and finetuning the remaining networks achieves the best AUC performance 0.7580, which is 0.0056 absolute gain over baseline model. Comparing different freeze-finetune split (methods 5-7), we can see that freezing pretrained parameters in level $L_{0}$ and $L_{1}$ (method 7) achieves better performance than other split methods (method 5/6), which is 0.0043 absolute gain in AUC. As for the ablation studies of Aspect Gating Fusion in table \ref{tab:aspect_gating_fusion}, the baseline is to simply mean pooling three aspects: ID, text and sparse features. We can see the Domain-Gating achieves best AUC performance 0.7524, which is 0.0139 absolute gain over mean-pooling method.


\subsubsection{Supervised Finetuning in Cold Start Scenarios}
To prove the effectiveness of finetuning our pretrained S\&R Foundation model, we compared cold start performance of two scenarios, Service Card Recommendation (task 6) and Content Query Recommendation (task 4). They are new scenarios and we only collected a few samples in a short period of time. The samples are splitted as we leave out last one day's collected data for testing, and use the remaining data for fine-tuning the S\&R Foundation. We also train the single domain model as the baseline. From table \ref{tab:supervise_finetune}, we can see the fine-tuned S\&R Foundation model achieves +0.0216 AUC improvement over single domain model on task 6 and +0.0279 AUC improvement on task 4. 

\subsubsection{Online AB Testing}
To further prove the effectiveness of online performance in cold start scenario, we deployed the fine-tuned S\&R Foundation model online in Service Card Recommendation scenario, and compared with baseline, which is the single domain DNN model. The results of the AB Testing from day 1 to day 7 are depicted in Figure \ref{fig:service_card_ab_testing}. The key performance measurement of the cold start scenario is PVCTR (Page View Click Through Rate). And we observed that the fine-tuned S\&R Foundation model achieved +17.54\% relative gain in PVCTR over baseline. The online AB Testing results showed that our method achieved better performance than baseline consistently in cold start scenario.

%% file: 6conclusion.tex
\vspace{-0.5em}
\section{Conclusion}
In this paper, we study the problem of training search and recommendation tasks jointly as the S\&R Multi-Domain Foundation model, and use domain adaptation techniques to benefit cold start scenario. Our proposed model learns user, query and item representations, applies LLM to encode domain invariant text features and Aspect Gating Fusion to merge ID, text and sparse features. We also conducted extensive experiments on finetuning the foundation models in cold start scenarios, which achieves better performance than the single domain model. The fine-tuned S\&R Multi-Domain Foundation model has been successfully deployed online in Alipay's multiple search and recommendation scenarios.